\documentstyle[prd,twocolumn,aps,epsf]{revtex}
\newcommand{\singlefig}[2]{
\begin{center}
\begin{minipage}{#1}
\epsfxsize=#1
\epsffile{#2}
\end{minipage}
\end{center}}
\begin{document}
\draft
\title{Gravitating BIon and BIon black hole with dilaton}
\author{Takashi Tamaki
\thanks{electronic
mail:tamaki@gravity.phys.waseda.ac.jp}}
\address{Department of Physics, Waseda University,
Ohkubo, Shinjuku, Tokyo 169-8555, Japan}
\author{Takashi Torii
\thanks{electronic mail:torii@resceu.s.u-tokyo.ac.jp}}
\address{Research Center for the Early Universe, 
University of Tokyo, Hongo, Bunkyo, Tokyo 113-0033, Japan
\\ and \\
Advanced Research Institute for Science and Engineering,
Waseda University, Ohkubo, 
Shinjuku, Tokyo 169-8555, Japan}
\date{\today}
\maketitle
%------------------------------
\begin{abstract}
We construct static and spherically symmetric particle-like 
and black hole solutions with magnetic or electric charge 
in the Einstein-Born-Infeld-dilaton system, which is a 
generalization of the Einstein-Maxwell-dilaton (EMD) system 
and of the Einstein-Born-Infeld (EBI) system. 
They have remarkable properties which are not seen for
the corresponding solutions in the EBI  
and the EMD system. In the electrically charged case, 
the electric field is neutralized at the origin by the effect of 
the diverging dilaton field. The extreme 
solution does not exist but we can take the zero horizon 
radius limit
for any Born-Infeld (BI) parameter $b$. 
Although the solution in this limit corresponds to 
the `particle-like' solution in the Einstein frame,
it is not a relevant solution in the string frame. 
In the magnetically charged case, the extreme solution {\it does}
exist for the critical BI parameter $\sqrt{b}Q_m=1/2$.
The critical BI parameter divides the solutions qualitatively.
For $\sqrt{b}Q_m<1/2$, there exists the particle-like solution
for which the dilaton field is finite everywhere, while
no particle-like solution exists and the solution in the
$r_h \to 0$ limit becomes naked for $\sqrt{b}Q_m>1/2$.
Examining the internal structure of the black holes,
we find that there is no inner horizon and that the
global structure is the same as the Schwarzschild one
in any charged case.
\end{abstract}

%%%%%%%%%%%%%%%%
%
%        Introduction
%
%%%%%%%%%%%%%%%%
The pioneering theory of the non-linear electromagnetic field was
formulated by Born and Infeld (BI) in 1934\cite{Born}.
Their basic motivation was to solve the problem of the self-energy 
of the electron by imposing a maximum strength for the electromagnetic
field. Although their attempt did not succeed in this regard, 
many kinds of solution, such as a vertex and a 
particle-like solution (BIon),
which were constructed afterward in the model including
BI term were of great interest.
Some extensions of the BI type action 
to the non-Abelian gauge field are
considered, although it is not determined uniquely
because of the ambiguity in taking 
the trace of internal space\cite{Tsey}. 
One of the interesting recent results is that
there exist classical glueball solutions which were prohibited in the 
standard Yang-Mills theory\cite{Gal}.

Moreover, it was considered to include the self-gravitational
effect of the isolated systems by putting the Einstein gravity
\cite{rapid}.
Then self-gravitating particle-like solutions (EBIon)\cite{foot2} 
and their black hole
solutions (EBIon black hole) were discovered analytically under the
static spherically symmetric ansatz\cite{Demi,Oliveira}.
The non-linearity of the electromagnetic field brings remarkable 
properties to avoid the black hole singularity problem
which may contradict the strong version of the Penrose cosmic
censorship conjecture in some cases. Actually a new non-linear
electromagnetism was proposed, which produces a nonsingular
exact black hole solution satisfying the weak energy 
condition\cite{Gar,Eloy}, and has distinct properties
from Bardeen black holes\cite{Bardeen}.
Unfortunately, the BI model is not this type.

Surprisingly it has been shown that the world volume
action of a D-brane is described by a kind of non-linear
BI action in the weak string coupling limit\cite{brane}.
In particular, the bosonic D-3 brane will 
have a world volume action that is the BI action.
In addition to this, BI action arises in the sting-generated
corrections if one considers the coupling of an
Abelian gauge field to the open bosonic string or
the open superstring\cite{Frad}. This system
also includes the dilaton and the antisymmetric Kalb-Ramond
tenser field, which we call the axion field.
Hence it is a direct extension of the Einstein-Maxwell-dilaton-axion
(EMDA) system, where the famous black hole solution was found with
the vanishing axion (i.e., in the EMD system)
by Gibbons and Maeda, and independently
by Garfinkle, Horowitz and Strominger (GM-GHS)\cite{GM-GHS}.
It is interesting that the non-linear electromagnetic field
appears in  string theory since one of the major
problems of sting theory is to cure the undesirable singularity
in general relativity.

In the earlier works on the solutions in the EBI system, 
the dilaton and the axion field are not taken into account
to the best of our knowledge.
We should not, however, 
ignore them from the point of view of string theory
(especially of the latter context)
since it is well known that
the GM-GHS solution has many different aspects
from the Reissner-Nortstr\"om (RN) black hole in the 
Einstein-Maxwell system. 
Our main purpose in this letter is to survey the particle-like 
and black hole solutions in the 
Einstein-Born-Infeld-dilaton(-axion) (EBID) system,
which we call dilatonic EBIon (DEBIon) and dilatonic
EBIon black hole (DEBIon black hole), respectively,
and to clarify the effect of the dilaton field.

%%%%%%%%%%%%%%%%
%
%        Model and Basic Equations
%
%%%%%%%%%%%%%%%%
{\it Model and Basic Equations}:
We start with the following action\cite{notes};
%------------------------------
\begin{eqnarray}
& & S=\int d^{4}x\sqrt{-g}\left[
\frac{R}{2\kappa^{2}}-\frac{  (\nabla\phi )^{2} }{\kappa^{2}}
-\frac{1}{24\kappa^{2}}e^{-4\gamma\phi}H^{2}
+L_{BI}\right],
\label{EBID}
\nonumber
\\ & & 
\end{eqnarray}
%------------------------------
where $\kappa^{2} := 8\pi $ and  $\gamma$ is the
coupling constant of the dilaton field $\phi$.
The three rank antisymmetric tenser field
$H^{2}= H_{\mu\nu\rho}H^{\mu\nu\rho}$ is 
expressed as, $H=dB+\frac{1}{4}A\wedge F$. 
$L_{BI}$ is the BI part of the Lagrangian which is written as
%------------------------------
\begin{eqnarray}
L_{BI}=\frac{b e^{2\gamma\phi}}{4\pi}
\left\{
1-\sqrt{1+\frac{e^{-4\gamma\phi}}{2b}P
-\frac{e^{-8\gamma\phi}}{16b^{2}}Q^{2}    }
\right\}  \ ,
\label{matter}
\end{eqnarray}
%------------------------------
where $P:= F_{\mu\nu}F^{\mu\nu}$ and 
$Q:= F_{\mu\nu}\tilde{F}^{\mu\nu}$.
A tilde denotes the Hodge dual.
Since we examine the electric and the magnetic monopole
cases separately, 
the term $Q$ vanishes and eventually
the axion field becomes trivial by putting the spherically
symmetric ansatz.
The dyon case including the axion field are summarized in
elsewhere\cite{Tamaki}. 
The BI parameter $b$ has physical interpretation of a 
critical field strength. In the string theoretical context,
$b$ is related to the inverse string tension 
$\alpha'$ by $b^{-1}=(2\pi \alpha ')^{2}$.
Notice that the action (\ref{EBID}) reduces to the EMD system in the 
limit $b \rightarrow \infty$ and
to the EBI system with the massless field for $\gamma =0$. 
Here, we 
concentrate on the case $\gamma =1$
which is predicted from superstring theory
and $\gamma=0$ for comparison.

We consider the metric of static and spherically symmetric,
%------------------------------
\begin{eqnarray}
ds^{2}=-f(r) e^{-2\delta (r)}dt^{2}  
+f(r)^{-1}dr^{2}+r^{2}d\Omega^{2} \ ,
\label{metric}
\end{eqnarray}
%------------------------------
where $f(r)=1-2m(r)/r$. 
The gauge potential has the following form: 
%------------------------------
\begin{eqnarray}
A&=&-\frac{w_{1}(r)}{r}dt-w_{2}(r)\cos \theta d\varphi.
\label{elemag}
\end{eqnarray}
%------------------------------
Since we do not consider the dyon solution, 
we drop the term $w_{2}$ (or $w_{1}$) in the electrically
(or magnetically) charged case. 
From the BI equation, we obtain that $w_{2}\equiv Q_{m}$ 
is constant and
%------------------------------
\begin{eqnarray}
\left(\frac{w_{1}}{r}\right)'
=-Q_{e}e^{2\phi-\delta}
\left(r^4+\frac{Q_{e}^{2}}{b}\right)^{-\frac12},
\label{ele2}
\end{eqnarray}
%------------------------------
where $Q_{e}:= w_{1}(\infty)$
and a prime denotes a derivative with respect to $r$. 
From this equation, the electric field $E_r=-(w_1/r)'$
does not diverge but takes the finite value at the origin.
The maximum value is $E_r=\sqrt{b}$
in the EBI system. As we will see, the electric field
vanishes at the origin in the EBID system
by the nontrivial behavior of $\phi$.
The potential $w_1$ is formally expressed as 
%------------------------------
\begin{eqnarray}
w_{1}
=-r\int^r_0 \frac{Q_{e}e^{2\phi-\delta}}{r^2}
\left(r^4+\frac{Q_{e}^{2}}{b}\right)^{-\frac12}dr,
\end{eqnarray}
%------------------------------
where we put $rw_1(0)=0$ without loss of generality.

By the above ans\"{a}ze the basic equations 
with $\gamma=1$ are written as follows.
%------------------------------
\begin{eqnarray}
m'=-U+\frac{r^2}{2} f(\phi')^{2} ,
\label{m1}  
\\
\delta'=-r(\phi')^{2} ,
\label{d1}   
\\
\phi^{\scriptstyle\prime\scriptstyle\prime}=-\frac{2}{r}\phi'
-\frac{2}{f}
\left[\left(\frac{m}{r}+U\right)\frac{\phi'}{r}-X \right] ,
\label{p1}
\end{eqnarray}
%------------------------------
where
%------------------------------
\begin{eqnarray}
U :=   e^{2\phi}\left(br^{2}
-\sqrt{b^2 r^4+bQ_e^2+bQ_m^2 e^{-4\phi}  }\right),
\label{BI}  
\\
X:=  be^{2\phi}
\left(\sqrt{\frac{br^4+Q_e^2}{br^4+Q_m^2e^{-4\phi}}}-1
\right).
\end{eqnarray}
%------------------------------
We drop the terms which do not vanish in the dyon
case but do in the monopole case.
Note that by introducing the dimensionless variables
$\bar{r}:= \sqrt{b}r$, $\bar{m}:= \sqrt{b}m$,
$\alpha_{e}:= \sqrt{b}Q_{e}$  and
$\alpha_{m}:= \sqrt{b}Q_{m}$,
we find that the parameters of the equation system are
$\alpha_{e}$ and $\alpha_{m}$.

As was already
pointed out by Gibbons and Rasheed\cite{Gibbons},
this system loses
electric-magnetic duality,
although the modification to satisfy the SL(2,R) duality
is also considered\cite{Rasheed}. Hence, when we 
obtained the black hole solution with
electric charge in the above action, we can not transform to 
the solution with
magnetic charge by duality. We are interested in 
the difference between
the black holes with the electric charge and those with the 
magnetic charge.

The boundary conditions at spatial infinity to satisfy the 
asymptotic flatness are
%------------------------------
\begin{eqnarray}
m(\infty)\equiv M=Const.,\ \delta (\infty)=0 ,\ \phi (\infty)=0\ .
\label{atinf}
\end{eqnarray}
%------------------------------
We also assume the existence of a regular 
event horizon at $r=r_{h}$ for the DEBIon black hole. 
So we have
%------------------------------
\begin{eqnarray}
m_{h}=\frac{r_h}{2},\;\; \delta_h< \infty ,
\;\;  \phi_h<\infty ,\;\;  
\phi'_{h}=\frac{2r_hX_h}{(1+2U_h)}.
\label{atEH2}
\end{eqnarray}
%------------------------------
The variables with subscript $h$ means that they are evaluated 
at the horizon.
Under these conditions, we obtained the black hole 
solution numerically.

%%%%%%%%%%%%%%%%
%
%        Electrically charged solution
%
%%%%%%%%%%%%%%%%
{\it Electrically charged solution}:
First, we investigate the electrically charged case.
Before proceeding to the DEBIon black hole, we briefly review
the solutions in the EBI system ($\gamma =0$).
In the $b\to \infty$ limit, the EBI system reduces to the EM system
and the RN black hole is the solution of the Eqs.~(\ref{m1})-(\ref{p1}).
For the finite $b$, we obtain EBIon and its black hole 
solutions analytically\cite{Demi,Oliveira}.
The metric functions
are written as $\delta \equiv 0$ and
%------------------------------
\begin{eqnarray}
m(r)&=&m_0+\frac{bQ_e^{2}}{3}\left[
\frac{r}{r^2+\sqrt{r^4+bQ_e^2}} \right.  
\nonumber \\
&& \left. +\frac{1}{\sqrt{b}Q_e}F\left(\frac{1}{\sqrt{2}},
\arccos \frac{\sqrt{b}Q_e -r^2}{\sqrt{b}Q_e +r^2}
\right)\right]  ,
\label{EBIm}
\end{eqnarray}
%------------------------------
where $F(k,\varphi)$ is
the elliptic function of the first kind. 
The constant $m_0$ is the mass of the central object. 
$m_0=0$ corresponds to the EBIon solution. 
As is seen from the metric function, the number of the horizon
depends on the parameter $\alpha_e$ and $m_0$ (or $M$).
We plot the $M$-$r_h$ relation in Fig.~1.
The solution branches are divided qualitatively by 
$\alpha_e=\alpha^{\ast}:=1/2$.
For $\alpha_e >\alpha^{\ast}$, 
there is a special value $M_0$
of which the analytic form is seen in Ref.~\cite{Oliveira}.
For $M<M_0$ the black hole and inner horizon
exist as the RN black hole while
only the black hole horizon exists for $M \geq M_0$.
The minimum mass solution in each branch corresponds to the
extreme solution. 
On the other hand, all the black hole
solutions have only one horizon and the global structure
is the same as the Schwarzschild black hole
for $\alpha_e <\alpha^{\ast}$.
In this case we can take the $r_h \to 0$ limit. These
solutions correspond to the EBIon solution with no horizon. 
Note however that,  since $m'(0) = \alpha_e$,
they have the conical 
singularity at the origin, which is the characteristic
feature of the self-gravitating BIons.
For $\alpha_e =\alpha^{\ast}$, the extreme solution is 
realized in the
$r_h \to 0$ limit. Although Demianski called it 
an electromagnetic geon \cite{Demi} which is regular everywhere,
it has a conical singularity like the other EBIons.

Next we turn to the $\gamma=1$ case.
In the $b\to \infty$ limit, i.e., in the EMD system,
the GM-GHS solution exists. 
The GM-GHS solution has three global charges, i.e.,
mass, the electric charge and the dilaton charge.
The last one depends on the former two, hence 
the dilaton charge is classified
into the secondary charge\cite{Bizon}. 
Although the GM-GHS solution has the electric charge as the 
RN solution, it does not have an inner horizon but has
spacelike singularity inside the event horizon
due to the effect of the dilaton field.
In the extreme limit, the event horizon coincides with the 
central singularity and a naked singularity appears at the 
origin\cite{GM-GHS}. There is no particle-like solutions in this
system.

For the finite value of $b$, we can not find the analytic
solution and use the numerical analysis. 
Since there is no non-trivial dilation configuration
in the $Q_e=0$ case, the dilaton hair is again the secondary
hair in this system.
We  plot the $M$-$r_h$ relation of 
the DEBIon black hole in Fig.~1.
We can find that all branches reach to $r_h=0$
in contrast to the EBIon case.
This is explained as follows.
The extremal solution has a degenerate horizon and
$2m'=1$ is realized on the horizon.
Hence, by Eq.~(\ref{m1})
%------------------------------
\begin{eqnarray}
\alpha_e^{ext}
=\sqrt{\frac{e^{-4\phi_h}}{4}+br_h^2 e^{-2\phi_h}}.
\label{degenerate}
\end{eqnarray}
%------------------------------
Since $\phi_h =0$
in the EBI system, the extreme solutions exist for 
$\alpha_e \geq \alpha^{\ast}$.
We will examine the behavior of the dilation field
around the horizon. On the horizon, the equation
of the dilaton field becomes
$f'_h\phi'_h =2X_h$.
Since $f'_h>0$ and $X$ is positive except for the origin 
in the electrically charged case, the dilaton field 
increases  around the horizon.
Next, we assume that there is an extremum point of
the dilaton field. At this point, the equation
of the dilaton field becomes
$f\phi^{\prime \prime} =2X$,
which implies that the extremum point must be a 
local minimum. 
From this behavior, we find that if $\phi_h\geq 0$ the 
dilaton field can not approach its asymptotic value 
$\phi(\infty)=0$. As a result, $\phi_h<0$.
This means that the value $\alpha_e$ for which the extreme 
solution exists must be larger than that in the EBI system.
Furthermore, $\phi_h$ diverges to minus infinity in the
$r_h\to 0$ limit as we will see below. As a result
$\alpha_e ^{ext} \to \infty$ and no extreme solution
exists for finite $\alpha_e$.

%%%%%%%%%%%%%%
% DEBIon solution
%%%%%%%%%%%%%%
As for the `particle-like' solution, which is not a relevent 
one we show below, we have to analyze
carefully. By performing Taylor expansion
for the basic equations around the origin to see the 
behavior of the dilaton fields, we find that the lowest 
order equation is inconsistent. This implies that the
dilaton field diverges at the origin.
Then we employ the new function $\psi :=e^{2\phi}$ 
and expand the field variables as
%------------------------------
\begin{eqnarray}
\psi = \sum_{\alpha, \; \beta}\psi_ {(\alpha, \beta)} 
r^{\alpha} (\ln r)^{\beta},
\;\;\;\;
m = \sum_{\gamma, \; \delta}m_ {(\gamma, \delta)} 
r^{\gamma} (\ln r)^{\delta}.
\label{expand}
\end{eqnarray}
%------------------------------
Substituting them into the basic equations and
evaluating the lowest order equations, we find
%------------------------------
\begin{eqnarray}
\phi \sim -\frac{1}{2} \ln(-4\sqrt{bQ_e^2}\ln r),
\;\;\;\;
m \sim -\frac{r}{4\ln r}.
\end{eqnarray}
%------------------------------
Note that the singular behavior of the mass function becomes
mild by a factor $1/\ln r$ and that the lapse function $\delta$
remains finite in spite of the divergence of the dilaton field.

The diverging behavior of the dilaton field is important
when we consider the solutions in the original string frame.
Since the conformal factor is $e^{-2\phi}$, the conformal 
transformation becomes singular at the origin,
and there is a possibility that such transformation 
gives pathological behavior. Actually, performing 
conformal transformation back to the string frame,
we find that the metric is degenerate at the origin,
and that the strong curvature singularity occurs. In this sense,
the `particle-like' solution we found is not a relevant one.

By integrating from the event horizon back to the origin
with suitable boundary conditions, we can examine the
internal structure of the black hole solutions. The dilaton
field monotonically decreases and diverges 
as $\approx \log r$. The electric field vanishes approaching
to the origin by the divergence of the dilaton field.
The mass function $m$  also diverges as  $\approx r^{-x}$,
$(0<x<1)$. Hence, the function $f$ does not have zero
except at the event horizon. As a result, the global structure
is Schwarzschild type.

%%%%%%%%%%%%%%
% temperature
%%%%%%%%%%%%%%
The temperature of the DEBIon black holes
is expressed by
%------------------------------
\begin{eqnarray}
T& =& \frac{e^{-\delta_h}}{4\pi r_h}(1+2U_h).
\label{temperature}
\end{eqnarray}
%------------------------------
We show them in Fig.~2.
The DEBIon black hole has
always a higher temperature than the GM-GHS solution 
($T=1/8\pi M$)
since the field strength of the gauge field becomes mild
by the BI field.
The specific heat of the DEBIon black hole is always negative while
the discontinuous change of the sign of the heat capacity occurs 
once or twice for the EBIon black holes depending on $\alpha_e$.
Since the EBIon black hole has an extreme limit for 
$\alpha_e>\alpha^{\ast}$,  the temperature becomes
zero in this limit.
On the other hand,
since there is no extremal solution for DEBIon
black holes,
their  temperature does not go to zero but diverges in the
$r_h \to 0$ limit. 
Hence the evolution by the Hawking
evaporation does not stop until 
the singular solution with $r_h \to 0$ when 
the surrounding matter field does not exist.

%%%%%%%%%%%%%%%%
%
%        Magnetically charged solution
%
%%%%%%%%%%%%%%%%
{\it Magnetically charged solution}:
Tuning now to the  magnetically charged case. 
EBIon and GM-GHS black holes with magnetic charge can be obtained 
from the following duality as $F \to \tilde{F}$, and 
$F \to e^{-2\phi}\tilde{F}$, $\phi \to -\phi$, respectively. 
So the relations $M$-$r_{h}$ and $M$-$1/T$ never change. 
As is noted above however, our DEBIon black hole has no such duality. 
So we can expect that the magnetically charged solutions have
different properties from electrically charged one.

From Eq. (\ref{m1}), 
%------------------------------
\begin{eqnarray}
\alpha_{m}^{ext}=\sqrt{\frac{1}{4}+br_{h}^{2}e^{2\phi_{h}}}
\label{ext}
\end{eqnarray}
%------------------------------
for the extreme solutions.
Thus, there is no extreme solution for $\alpha_m <1/2$. 
For $\alpha_m \geq 1/2$, 
we must survey $\phi_h$. 
For $\alpha_m =1/2$, because 
$r_{h}^{2}e^{2\phi_h} \to 0 $ as $r_h \to 0$, extreme solution is realized 
in the $r_h \to 0$ limit, i.e., $\alpha_{m}=\alpha_{m}^{ext}$. 
For $\alpha_m >1/2$, $r_{h}^{2}e^{2\phi_h} \to Const.$ as 
$r_h \to 0$. We cannot tell whether Eq.~(\ref{ext}) is fulfilled
for a certain horizon radius since $\phi_h$ is obtained
iteratively only by numerical method.
Our calculation always shows 
$\alpha_{m}<\sqrt{1/4+br_{h}^{2}e^{2\phi_{h}}}$
except for $\alpha_{m}=\infty$

Although the $M$-$r_h$ relation seems similar to that of  
the GM-GHS solution 
regardless of $\alpha_m$ (Fig. 3), the $M$-$1/T$ relation
is different depending on $\alpha_m$ (Fig. 4).
This is due  to the qualitative difference of the behavior
of the dilaton field in the $r_h \to 0$ limit. For 
$\alpha_m \geq\alpha^{\ast}$,
the dilaton field on the horizon diverges as
$\phi_h \approx \ln r_h$ and $e^{-\delta_h} \to r_h$, and
hence the temperature remains finite as in the GM-GHS case.
On the other hand, $\phi_h$ and $\delta_h$ are finite in the
$r_h \to 0$ limit for $\alpha_m <\alpha^{\ast}$. As a result,
the temperature diverges. Note that the corresponding solution
is the particle-like solution. 
The remarkable feature is that the dilaton field is finite
everywhere. Hence it is also the relevant particle-like solution in
the original string frame.
In the magnetically charged case,
DEBIon solution does exist for $\alpha_m <\alpha^{\ast}$.

The behaviors of the field functions inside the event
horizon are similar to those in the electrically charged
case qualitatively except for the sign of the dilaton
field. The magnetic field diverges as $B_r \approx
r^{-x}$, $(1<x<2)$.

%%%%%%%%%%%%%%%%
%
%        Discussion
%
%%%%%%%%%%%%%%%%
{\it Discussion}: 
We investigate the static spherically symmetric solutions
in the EBID system. In the electrically charged case, there
is neither extreme solution nor particle-like solution both of which
exist in the EBI system. The temperature
diverges monotonically in the $r_h \to 0$ limit.
On the other hand, the extreme solution exists when 
$\sqrt{b}Q_m =1/2$ and particle-like solutions do when 
$\sqrt{b}Q_m <1/2$ for the magnetically charged case.
For $\sqrt{b}Q_m >1/2$, the solution in the $r_h \to 0$ limit
corresponds to naked singularity. The behavior
of the temperatures varies 
depending on $\sqrt{b}Q_m$. We do not find
the inner horizon in any charged case, and
the global structure is the Schwarzschild type.
This is due to the effect of the dilaton field.

Here, we discuss some outstanding issues. 
One of the most important issues is the stability of the new
solutions. Here we briefly discuss this by using 
catastrophe theory\cite{cata,torii}, 
which is a useful
tool to investigate relative stability. Adopting $M$, $Q_e$,
(or $Q_m$), $b$, $F^2_h$ and $r_h$ as catastrophe variables,
we find that the equilibrium space becomes single 
valued space with
respect to the control parameters and includes the 
stable GM-GHS solution. This implies that DEBIon black hole
is also stable at least against spherical perturbations.

We also discuss the possibility of  
relating inverse string tension $\alpha'$ and the gravitational 
constant $G$. Following the lines in Ref. \cite{Gibbons}, 
the supersymmetric spin $1/2$, $0$ particle is described by 
BI action if we choose 
%------------------------------
\begin{eqnarray}
\frac{1}{b}=\frac{2}{3}\left(\frac{e}{m}\right)^{4}, 
\end{eqnarray}
%------------------------------
where $e$ and $m$ is the U(1) charge and the mass of the particle, 
respectively. Identifying the particle with extreme black hole 
solution and approximating it to the RN solution, we obtain 
$2\pi \alpha'\sim \sqrt{2/3}G$.  
If we apply this discussion to the system including 
BI action, it may be suitable using $e/m$ ratio 
of the EBIon black hole and DEBIon black holes with 
$\alpha_{e}$(or $\alpha_{m}$)$=\alpha^{\ast}$, in which cases, 
extreme solution is realized in the zero horizon limit. 
Then, we find $2\pi \alpha'\sim 1.07G$ for EBIon black hole,  
and $2\pi \alpha'\sim 1.73G$ for DEBIon black hole.  
So the discussion in \cite{Gibbons} is not much affected.  
More detailed properties and the dyon case including 
the axion are shown 
in our next paper\cite{Tamaki}.

%%%%%%%%%%%%%%%%
%
%        Acknowledgments
%
%%%%%%%%%%%%%%%%
{\it Acknowledgments}:
Special thanks to Gary W. Gibbons and Kei-ichi Maeda for useful
discussions.  T. T and T. T are thankful for  financial support from the
JSPS. This work was supported  by a JSPS Grant-in-Aid
(No. 094162 and No. 106613), and 
by the Waseda University Grant  for Special
Research Projects.

%%%%%%%%%%%%%%%%%%%%%%%%%%%%%%%%%%%%%%%

%%%%%%%%%%%%%%%%%%%
\begin{figure}
\begin{center}
\singlefig{8cm}{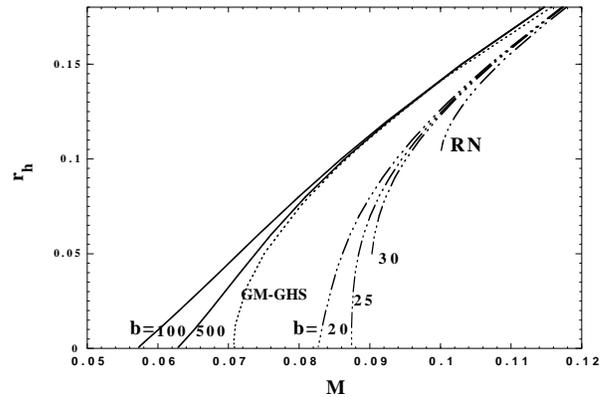}
\caption{$M$-$r_h$ diagram of the electrically charged
DEBIon black holes (solid lines) with 
electric charge $Q_e =0.1$ and $b=100$ and $500$. 
GM-GHS and EBIon black holes are 
also plotted by dotted line and the dot-dashed lines, respectively. 
\label{M-rhele} }
\end{center}
\end{figure}
%%%%%%%%%%%%%%%%%%%%
\begin{figure}
\begin{center}
\singlefig{8cm}{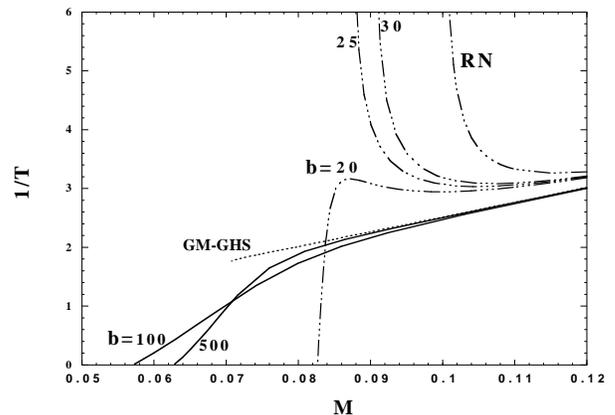}
\caption{$M$-$1/T$ diagram of the electrically charged
DEBIon black holes (solid lines) 
with the same parameters in Fig.~1.  
\label{M-1Tele} }
\end{center}
\end{figure}
%%%%%%%%%%%%%%%%%%%%
\begin{figure}
\begin{center}
\singlefig{8cm}{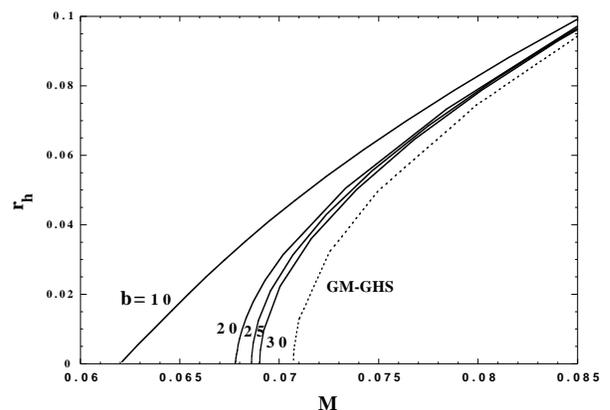}
\caption{$M$-$r_h$ diagram of the magnetically charged
DEBIon black holes (solid lines)  with
magnetic charge $Q_m =0.1$ and $b=10$, $20$, $25$ and $30$. 
GM-GHS black hole is 
also plotted by dotted line. 
\label{M-rhmag} }
\end{center}
\end{figure}
%%%%%%%%%%%%%%%%%%%%
\begin{figure}
\begin{center}
\singlefig{8cm}{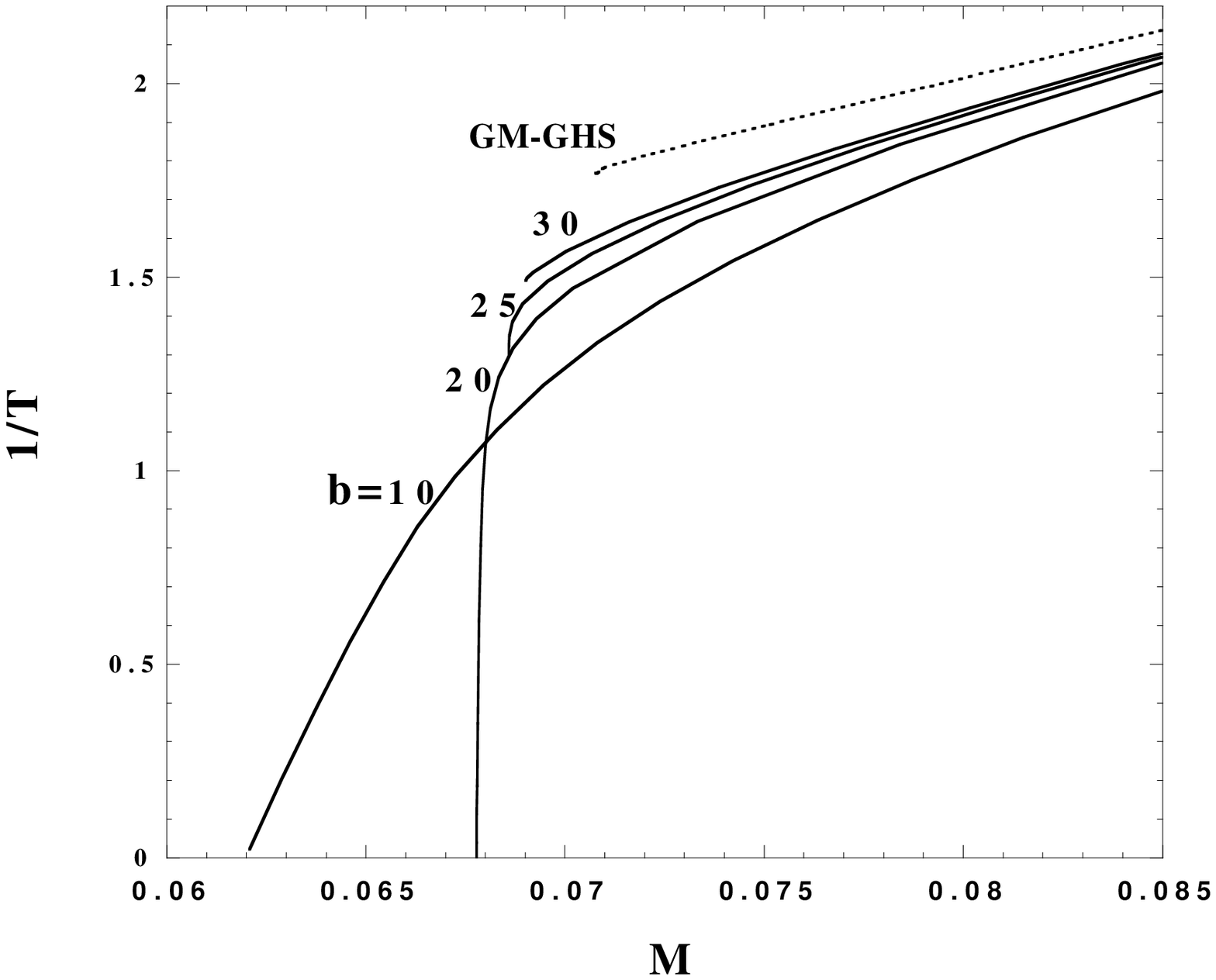}
\caption{$M$-$1/T$ diagram of the magnetically charged
DEBIon black holes (solid lines)
with the same parameters in Fig. 3. 
\label{M-1Tmag} }
\end{center}
\end{figure}
%%%%%%%%%%%%%%%%%%%%

\begin{thebibliography}{99}
\bibitem{rapid}
Black hole solutions in the SU(2) 
EBI system are also reported recently\cite{Kunz}.
\bibitem{foot2}
In this word we mean the solution which has conical singularity 
at the origin and has finite energy.
\bibitem{notes}
In this paper, we study the tree-level action. When we
are interested in the higher-loop action, the coupling constant $\gamma$
takes different value by assuming anomaly cancellation 
(See. Ref.~\cite{Callan}).
\end{thebibliography}
\end{document}